\pdfoutput=1
\documentclass[12 pt]{article}
\usepackage{graphicx}
\usepackage{amssymb}
\usepackage{amsmath}
\usepackage{amsfonts}
\usepackage{amsthm}
\usepackage[english]{babel}
\usepackage[latin1,ansinew]{inputenc}
\usepackage{a4wide}
\usepackage{color}
\newcommand{\be}{\begin{eqnarray}}
\newcommand{\ee}{\end{eqnarray}}

\begin{document}
\date{}
\title{\textbf{On the Jump Conditions for Shock Waves in Condensed Materials}}
\author{\textbf{Raj Kumar Anand} \\Department of Physics, UGC Centre of Advanced Studies,
\\University of Allahabad, Prayagraj 211002, India
\\ \textit{email: rkanand@allduniv.ac.in, anand.rajkumar@rediffmail.com}}




\maketitle

%

\begin{abstract}
In this article, we have proposed Rankine-Hugoniot (RH) boundary conditions at the normal shock front, which is passing through the condensed material. These RH conditions are quite general, and their convenient forms for the particle velocity, mass density, pressure, and temperature have been presented in terms of the upstream Mach number and the material parameters for the weak and the strong shocks, respectively. Finally, the effects on the mechanical quantities of the shock-compressed materials, e.g., titanium Ti6Al4V, stainless steel 304, aluminum 6061-T6, etc., have been discussed.
\end{abstract}

\section{Introduction}
\label{intro}
The Rankine-Hugoniot (RH) boundary conditions are found taking into account the conservation laws of mass, momentum, and energy across a shock front occurring in a compressible medium. The particle velocity, mass density, pressure, temperature, and flow speed show abrupt change or discontinuity across the shock front. An outstanding review has been written by Krehl \cite{Krehl2015} on the RH jump conditions at a shock in compressible gases. The high pressures (from 10$^{2}$ MPa to 10$^{3}$ GPa) and temperatures (up to 10$^4$ K) produced in the shock-compressed solids may change their properties, such as crystal structure, melting, and vaporization. Under such extreme pressures and temperatures, the behaviour of solid materials may be assumed to be like fluids, and, therefore, the laws of compressible fluids may be applied to the shock-compressed solid materials \cite{Kanel2004}.  Therefore, the study of shock compressed-solids, e.g., stainless steel, tungsten, copper, etc., and superhard materials, e.g., diamond, boron nitrides, etc., is essential to disclose their behaviour. 

An equation of state for solid materials was proposed by Mie \cite{Mie1903} and Gr\"uneisen \cite{Gruneisen1912}, which is known as the Mie-Gr\"uneisen equation of state (MG-EOS). To investigate the behaviour of shock-compressed materials Bushman and Fortov \cite{Bushman1983}, Anisimov and Kravchenko \cite{Anisimov1985}, Lieberthal et al. \cite{Lieberthal2017}, Ramsey et al. \cite{Ramsey2018}, Anand and Singh \cite{Anand2024}, and many others have used the following form of MG-EOS: $p=\Gamma e\rho$, where $p$ is the pressure, $e$ is the specific internal energy, $\rho$  is the mass density, and $\Gamma$ is the Gr\"uneisen coefficient. Obviously, in the case of an ideal-gas EOS, $\Gamma=(\gamma-1)=$ constant, where $\gamma$ is an adiabatic index. In terms of bulk modulus $K$  and thermal expansion coefficient $\alpha$, the Gr\"uneisen coefficient may be written as $\Gamma=v (\frac{dp}{de})_v = v\alpha K/c_v$, where $v$ is the specific volume, and $c_v$ is the specific heat at constant volume. Bushman and Fortov \cite{Bushman1983} have studied the shocked condensed materials taking into account  $\Gamma(G)=\frac{2}{3}+\left(\Gamma_o-\frac{2}{3}\right)\frac{G_m^2+1}{G_m^2+G^2}G$. It is notable that the material parameters $\Gamma_o$ and $G_m$ are generally determined by experiments, and $G(=\rho/\rho_o)$ is the shock compression ratio. The MG-EOS has correct asymptotics at $G\rightarrow 0$ and $G\rightarrow \infty$ , and it describes qualitatively the thermal components of pressure in a wide pressure range \cite{Anisimov1985} . The local speed of sound $a$ in condensed material is given by $a=\sqrt{(\Gamma+1)p/\rho}=\sqrt{K_s /\rho}$, where $K_s$ is the adiabatic bulk modulus, which provides information about the thermodynamic properties of the material.

This article presents the RH boundary conditions across a normal shock wave of finite thickness propagating in a shock-condensed material, taking into account the following assumptions: (i) The condensed material follows MG-EOS and is homogeneous, isotropic, and chemically nonreactive; (ii) the viscosity and thermal conductivity of the material are negligible; and (iii) the dissociation and ionization of molecules are very small. The RH boundary conditions for the particle velocity, mass density, pressure, and temperature have been written in terms of the Gr\"uneisen coefficient $\Gamma$ and the upstream shock Mach number $M$. Further, the suitable forms of the RH boundary conditions, respectively, across the weak and strong shock fronts have also been proposed in terms of the material parameters $\Gamma_o$ and $G_m$. Finally, the effects on the mechanical quantities of the shock-compressed aluminum 6061-T6, OFHC copper, titanium Ti6Al4V, brass (66\% copper and 34\% zinc), stainless steel 304, tantalum, and iron have been investigated. Thus, this study provided a clear picture of whether and how the mechanical quantities differ for the shock-compressed materials.  

The structure of this paper is organized as follows: In the next Sect. \ref{conditions}, we present the construction of RH shock jump conditions for condensed materials. Section \ref{results} contains the analysis with discussion on important components. The concluding remarks are given in Sect. \ref{concl}. 

\section{RH shock jump conditions}
\label{conditions}
In this section, RH jump conditions have been proposed for the weak and strong shock waves propagating in the shock-compressed solid materials. Let us now consider a shock front in a one-dimensional unsteady fluid. If a scalar conservation law is written in the integral form
$\oint _c Pdr-Qdt=0$, where $P$ and $Q$ are functions of $r$ and $t$, and $c$ is any smooth closed curve in the region in which the solution is required, then a differentiable solution will satisfy the partial differential equation $\frac{\partial P}{\partial t}+\frac{\partial Q}{\partial r}=0$, and a jump condition across a shock, i.e., at $r=R(t)$, is $\frac{dr}{dt}=\frac{[Q]}{[P]}$. 
We can write the integral forms of the conservation laws of mass, momentum, and energy as: $\oint _c \rho dr-\rho u dt=0$, $\oint _c \rho udr-(\rho u^2+p) dt=0$ and $\oint _c (\rho u^2/2 +\rho e) dr-(\rho u^3/2 + \rho eu + pu) dt=0$, respectively, where $e=c_vT$ is the internal energy per unit mass, and $T$ is the temperature.
Using MG-EOS, the above conservation laws lead to the shock conditions: $U=\frac{[\rho u]}{\rho}$, $U=\frac{[\rho u^2+p]}{\rho u}$, and  $U = \frac{[\rho u^3/2+(\Gamma+1)pu/\Gamma]}{\rho u^2/2+p/\Gamma}$, where $u$ is the particle velocity, and $U=\frac{dR}{dt}$ is the shock velocity. These conditions are called RH boundary conditions or RH shock jump relations. We can easily write these shock conditions after some algebraic manipulations as: $\left[\rho(u-U)=0\right]$, $\left[p+\rho(u-U)^2=0\right]$, and $\left[p(\Gamma+1)/\rho \Gamma + (u-U)^2/2\right]=0$. Fluid-mechanical quantities in the shock-compressed and uncompressed states of the material are related by the RH boundary conditions.

Now, let us consider a shock front passing through a motionless, homogeneous, condensed material having constant initial mass density $\rho_o$. In an Eulerian (laboratory) frame, at an equilibrium state $(u=u_o=0, p=p_o, \rho=\rho_o)$, the shock conditions can be written as:
\begin{eqnarray}\label{eq-1}
\rho(U-u)&=&\rho_o U,\nonumber \\
p+\rho(U-u)^2&=&p_o+\rho_o U^2,\\
\frac{p(\Gamma+1)}{\rho\Gamma}+\frac{(U-u)^2}{2}&=&\frac{p_o(\Gamma+1)}{\rho_o\Gamma}+\frac{U^2}{2}.\nonumber
\end{eqnarray}
where the quantities in shock-compressed material are with subscript $o$,  and the quantities in unshocked material are without subscript.
 The Mach number of the shock front is given by $M_s=U/a_o$, where $a_o= \sqrt{(\Gamma_o+1)p_o/\rho_o}$ is the initial speed of sound in condensed material. In further analysis, $M_s$ will be written without subscript $s$, i.e., $M_s \approx M$ (say). Let us now establish a relation between the upstream Mach number $M$ and the downstream Mach number $M_f[=(u-U)/a]$. Using some basic algebraic operations, $M_f$ may be written in terms of the experimentally measured $M$ as:
\begin{equation}\label{eq-2}
M_f^2=\frac{(\Gamma_o+1)(2+\Gamma M^2)}{(\Gamma+1)[2M^2(\Gamma+1)-\Gamma]}.
\end{equation}

\begin{figure}   
   \begin{minipage}{0.4\textwidth}
     \centering
     \includegraphics[width=1.4\linewidth, trim=0 .1cm .1cm 0, clip]{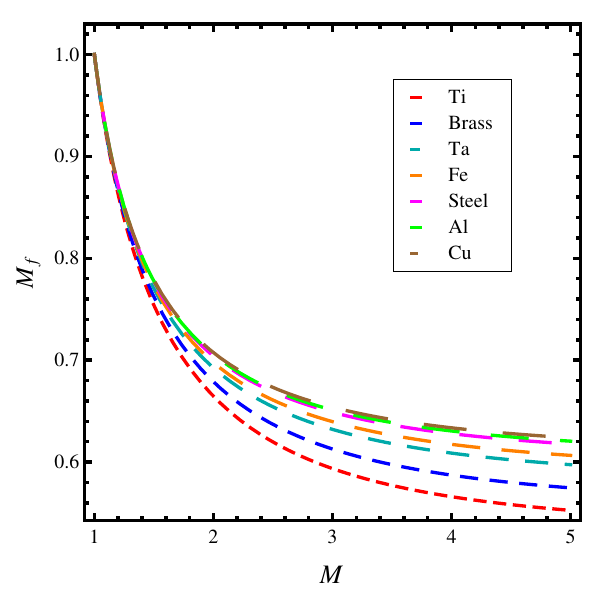}
       \end{minipage}\hfill
       \caption{Graphs of $M_f$ vs. $M$  with $G_m$.}\label{fig_1}
\end{figure}

Now, there are three unknowns $(\rho, u, p)$ in the set of three equations (\ref{eq-1}), so the remainder of the effort to relate the downstream and upstream conditions at the shock front is primarily algebra. Elimination of $\rho$ and $u$ from the set of equations (\ref{eq-1}) gives the ratios of pressure, density, and particle velocity. These relations are: 
\begin{eqnarray}\label{eq-3}
\frac{p}{p_{o}}&=&\frac{2\left(\Gamma+1\right)M^{2}}{\Gamma+2}-\frac{\Gamma}{\Gamma+2},\nonumber \\
\frac{\rho}{\rho_{o}}&=&\frac{\left(\Gamma+2\right)M^{2}}{\Gamma M^{2}+2},\\
\frac{u}{a_{o}}&=&\frac{2}{\Gamma+2}\left(M-\frac{1}{M}\right).\nonumber 
\end{eqnarray}
Here $p/p_o$ and $\rho/\rho_o$ are increasing functions of the shock Mach number $M$. The case $M=1$ is trivial as it corresponds to $p=p_o$ and $\rho=\rho_o$. The other two cases are (i) the rarefaction shock $(M<1, M_f>1, p<p_o, \rho<\rho_o)$ and (ii) the compression shock $(M>1, M_f<1, p>p_o, \rho>\rho_o)$. The fact that $M>1$ while $M_f<1$ means that the shock travels supersonically relative to the unshocked material and subsonically relative to the shocked material. Now, using Eq.(\ref{eq-3}), the dimensionless expressions for the temperature, the speed of sound, the adiabatic bulk modulus, and the change in entropy of the material can be, respectively, expressed as:
\begin{eqnarray}\label{eq-4}
\frac{T}{T_{o}}&=&\frac{\left[2\left(\Gamma+1\right)M^{2}-\Gamma\right](2+\Gamma M^{2})}{\Gamma\left(\Gamma +2\right)^{2} M^{2}},\nonumber \\
\frac{a}{a_{o}}&=&\left(\frac{\left(\Gamma+1\right)\left(2+\Gamma M^{2}\right)\left[2M^{2}\left(\Gamma+1\right)-\Gamma\right]}{(\Gamma_{o}+1)(\Gamma+2)^{2}M^{2}}\right)^{1/2},\\
\frac{K_s}{p_{o}}&=&\frac{(\Gamma+1)[2(\Gamma+1)M^2-\Gamma]}{\Gamma+2},\nonumber\\
\frac{\Delta s}{c_v}&=&ln \frac{2(\Gamma+1)M^2-\Gamma}{\Gamma+2} - (\Gamma+1)ln \frac{(\Gamma+2)M^2}{\Gamma M^2+2}.\nonumber
\end{eqnarray}
Now, we insist that the shock be compressive so that $p\geq p_o$. This inequality is satisfied in practice and indicates that $M\geq1$, $M_f\leq1$, $\rho\geq\rho_o$, $u\geq a_o$, $T\geq T_o$, and $a\geq a_o$. Further,the inequality $p\geq p_o$ also suggests that $p/\rho^{\Gamma+1}\geq p_o/\rho_o^{\Gamma+1}$ which means that the entropy change in matters due to the passage of shock is negative for rarefaction shocks and positive for compression shocks. Therefore, only compression shocks are physically feasible, whereas rarefaction shocks are not possible. This is also supported by the second law of thermodynamics. It is notable that all shocks involve dissipation and irreversibility. 

\section{Results and Discussion}
\label{results}
This section presents an exploration of RH conditions across a normal shock front propagating in the condensed materials. OFHC copper, aluminum 6061-T6, titanium Ti6Al4V, stainless steel 304, brass (66\% copper and 34\% zinc), tantalum, and iron have potential applications in military, industries, aerospace and automobiles. The numerical values of Gr\"uneisen parameter for such materials are given in Table \ref{tab:1}. The range of numerical values of $G_m$ lies between 0.5 and 0.8 \cite{Anisimov1985}. 

\begin{table}
\caption{Gr\"uneisen parameter $\Gamma_o$ (see Refs.\cite{Bushman1983,Anisimov1985,Steinberg1996})}
\label{tab:1}
\begin{tabular}
{p{0.6 cm} p{1.8 cm} p{1.8 cm} p{1.8 cm} p{1.8 cm} p{1.8 cm} p{1.8 cm} p{1.8 cm} p{1.8 cm}}
\hline\noalign{\smallskip}
 & Titanum & Brass & Tantalum & Iron & Stainless & Aluminum    & OFHC   \\
 & Ti6Al4V & &   &  & steel 304 & 6061-T6  & copper \\
 \noalign{\smallskip}\hline\noalign{\smallskip}
 $\Gamma_o$ & 1.23 & 1.43 & 1.67 & 1.78 & 1.93 & 1.97 & 2.02 \\
  \noalign{\smallskip}\hline
\end{tabular}
\end{table}

The value of shock compression ratio $G$ is found from the solution of an equation that is obtained using Eq. (\ref{eq-2}) and the second relation of Eq. (\ref{eq-3}). Thus, taking $M=3$ and using Table \ref{tab:1}, the calculated values of $G$ for titanium Ti6Al4V, brass (66\% copper and 34\% zinc), tantalum, iron, stainless steel 304, aluminum 6061-T6, and OFHC copper are given in Table \ref{tab:2}.  

\begin{table}
\caption{Computed values of material parameter $G$}
\label{tab:2}
\begin{tabular}
{p{1.2 cm} p{1.8 cm} p{1.8 cm} p{1.8 cm} p{1.8 cm} p{1.8 cm} p{1.8 cm} p{1.8 cm} p{1.8 cm}}
\hline\noalign{\smallskip}
$G_{m}$ & Ti & Brass & Ta  &  Fe & Steel & Al  & Cu \\
 \noalign{\smallskip}\hline\noalign{\smallskip}
 0.51 & 2.53522 & 2.39671 & 2.24879 & 2.18742 & 2.10990 & 2.09039 & 2.06665 \\
 0.52 & 2.53251 & 2.39348 & 2.24520 & 2.18375 & 2.10619 & 2.08667 & 2.06294 \\
 0.53 & 2.52976 & 2.39021 & 2.24158 & 2.18005 & 2.10245 & 2.08293 & 2.05920 \\
 0.54 & 2.52697 & 2.38690 & 2.23793 & 2.17632 & 2.09867 & 2.07916 & 2.05544 \\
 0.55 & 2.52416 & 2.38356 & 2.23424 & 2.17255 & 2.09488 & 2.07537 & 2.05166 \\
 0.56 & 2.52130 & 2.38019 & 2.23052 & 2.16876 & 2.09106 & 2.07155 & 2.04785 \\
 0.57 & 2.51842 & 2.37678 & 2.22677 & 2.16495 & 2.08722 & 2.06772 & 2.04403 \\
 0.58 & 2.51550 & 2.37334 & 2.22300 & 2.16111 & 2.08336 & 2.06387 & 2.04019 \\
 0.59 & 2.51256 & 2.36987 & 2.21920 & 2.15725 & 2.07949 & 2.06000 & 2.03634 \\
 0.60 & 2.50958 & 2.36637 & 2.21538 & 2.15337 & 2.07560 & 2.05611 & 2.03247 \\
 0.61 & 2.50657 & 2.36285 & 2.21154 & 2.14947 & 2.07169 & 2.05222 & 2.02860 \\
 0.62 & 2.50353 & 2.35929 & 2.20767 & 2.14555 & 2.06778 & 2.04832 & 2.02471 \\
 0.63 & 2.50047 & 2.35572 & 2.20379 & 2.14162 & 2.06385 & 2.04440 & 2.02082 \\
 0.64 & 2.49738 & 2.35211 & 2.19989 & 2.13768 & 2.05991 & 2.04048 & 2.01692 \\
 0.65 & 2.49427 & 2.34849 & 2.19597 & 2.13372 & 2.05597 & 2.03655 & 2.01302 \\
 0.66 & 2.49113 & 2.34484 & 2.19204 & 2.12976 & 2.05203 & 2.03263 & 2.00912 \\
 0.67 & 2.48797 & 2.34118 & 2.18810 & 2.12578 & 2.04808 & 2.02869 & 2.00522 \\
 0.68 & 2.48478 & 2.33749 & 2.18415 & 2.12180 & 2.04413 & 2.02476 & 2.00131 \\
 0.69 & 2.48157 & 2.33379 & 2.18019 & 2.11781 & 2.04018 & 2.02083 & 1.99741 \\
 0.70 & 2.47834 & 2.33007 & 2.17622 & 2.11382 & 2.03623 & 2.01691 & 1.99352 \\
 0.71 & 2.47510 & 2.32633 & 2.17224 & 2.10983 & 2.03228 & 2.01298 & 1.98963 \\
 0.72 & 2.47183 & 2.32258 & 2.16826 & 2.10584 & 2.02834 & 2.00907 & 1.98575 \\
 0.73 & 2.46854 & 2.31882 & 2.16428 & 2.10185 & 2.02441 & 2.00516 & 1.98188 \\
 0.74 & 2.46524 & 2.31505 & 2.16029 & 2.09786 & 2.02048 & 2.00126 & 1.97802 \\
 0.75 & 2.46192 & 2.31127 & 2.15631 & 2.09387 & 2.01656 & 1.99737 & 1.97417 \\
 0.76 & 2.45859 & 2.30748 & 2.15232 & 2.08989 & 2.01265 & 1.99349 & 1.97033 \\
 0.77 & 2.45524 & 2.30368 & 2.14834 & 2.08592 & 2.00876 & 1.98962 & 1.96651 \\
 0.78 & 2.45188 & 2.29988 & 2.14436 & 2.08195 & 2.00487 & 1.98577 & 1.96270 \\
 0.79 & 2.44851 & 2.29607 & 2.14038 & 2.07799 & 2.00101 & 1.98194 & 1.95891 \\
 \noalign{\smallskip}\hline
\end{tabular}
\end{table}

\begin{figure}   
   \begin{minipage}{0.4\textwidth}
     \centering
     \includegraphics[width=1.1\linewidth, trim=0 .1cm .1cm 0, clip]{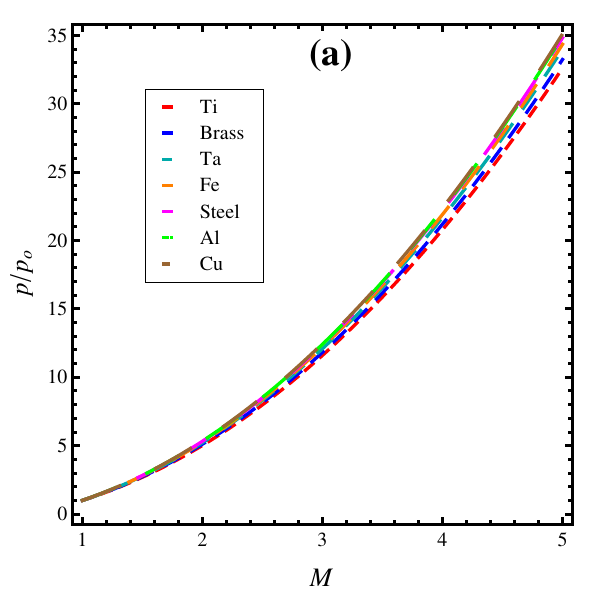}
       \end{minipage}\hfill
   \begin{minipage}{0.4\textwidth}
     \centering
     \includegraphics[width=1.10\linewidth, trim=0 .1cm .1cm 0, clip]{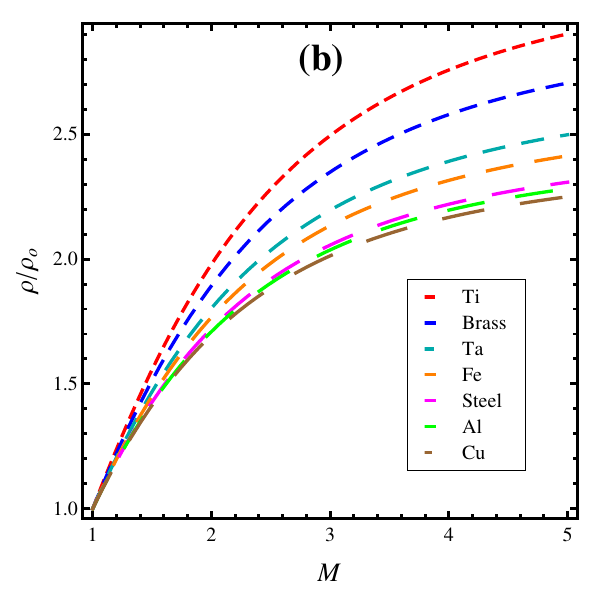}
     \end{minipage}
     \begin{minipage}{0.4\textwidth}
     \centering
     \includegraphics[width=1.1\linewidth, trim=0 .1cm .1cm 0, clip]{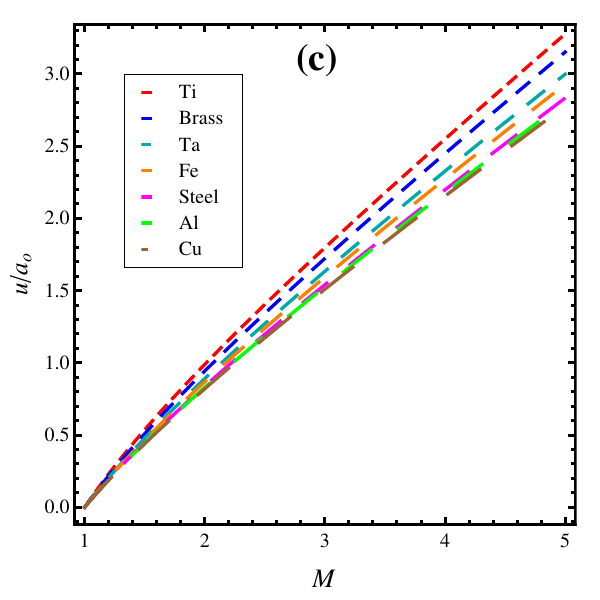}
       \end{minipage}\hfill
   \begin{minipage}{0.4\textwidth}
     \centering
     \includegraphics[width=1.10\linewidth, trim=0 .1cm .1cm 0, clip]{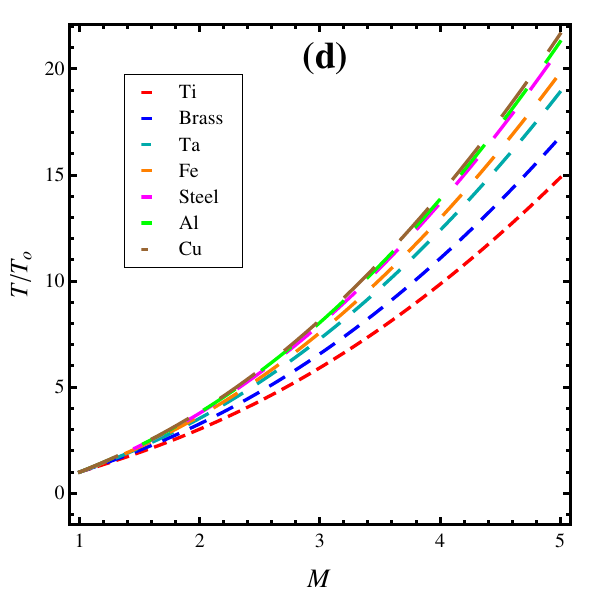}
     \end{minipage}
     \begin{minipage}{0.4\textwidth}
     \centering
     \includegraphics[width=1.1\linewidth, trim=0 .1cm .1cm 0, clip]{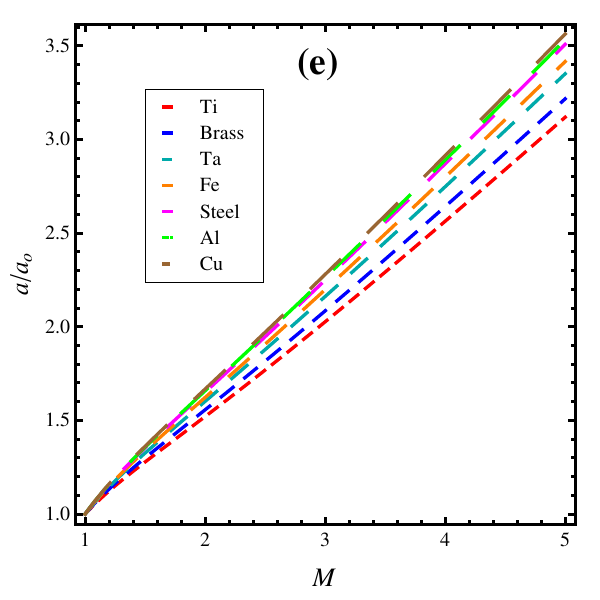}
       \end{minipage}\hfill
   \begin{minipage}{0.4\textwidth}
     \centering
     \includegraphics[width=1.1\linewidth, trim=0 .1cm .1cm 0, clip]{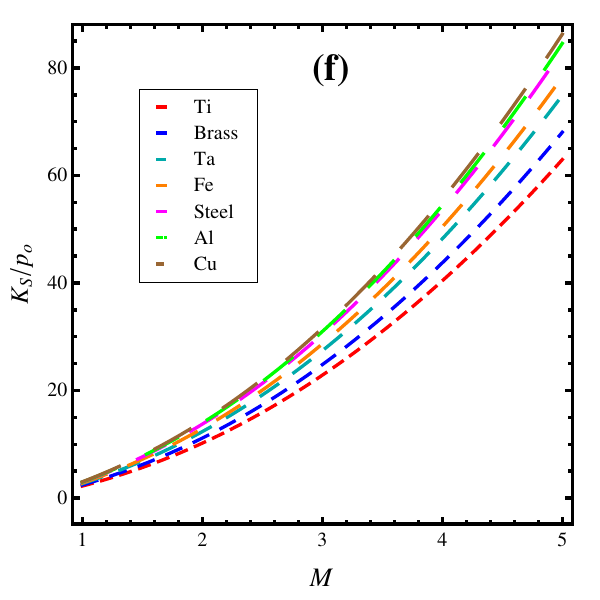}
     \end{minipage}
     \caption{Graphs of (a) $p/p_{o}$ vs. $M$, (b) $\rho/\rho{o}$ vs. $M$, (c) $u/a_o$ vs. $M$, (d) $T/T_o$ vs. $M$, (e) $a/a_o$ vs. $M$, (f) $K_s/p_o$ vs. $M$ with $G_m$=0.65.}\label{fig_2}
\end{figure}

Figure \ref{fig_1} depicts the variation of downstream Mach number $M_f$ versus upstream Mach number $M$ with $G_m$= 0.65. It shows that the downstream Mach number $M_f$ decreases with an increase in the upstream Mach number $M$. It is noticeable that when $M>1$, then Eq. (\ref{eq-2}) requires that $M_f<1$. Thus, $M_f$  changes from supersonic to subsonic values across a shock front, and this is the only possibility.  The downstream Mach number $M_f$ decreases with increasing values of Gr\"uneisen parameter $\Gamma_o$, especially when $M>1$. Obviously, the value of $M_f$ is maximum for OFHC copper and minimum for titanium Ti6Al4V.

The variations of the pressure $p/p_o$, the mass density $\rho/\rho_o$, the particle velocity $u/a_o$, the temperature $T/T_o$, the speed of sound $a/a_o$, and the adiabatic bulk modulus $K_s/p_o$ versus the upstream Mach number $M$ with $G_m$ =0.65 are shown in Fig. \ref{fig_2}. Figure \ref{fig_2} shows that in titanium Ti6Al4V, brass (66\% copper and 34\% zinc), tantalum, iron, stainless steel 304, aluminum 6061-T6 and OFHC copper, the pressure, the mass density, the particle velocity, the temperature, the speed of sound, and the adiabatic bulk modulus increase with an increase in the upstream Mach number $M$. The mass density increases rapidly for the values $M\leq2.5$ and then it increases slowly. The variations in the pressure, the temperature, the speed of sound, and the adiabatic bulk modulus are maximum in the shocked OFHC copper, whereas these variations are minimum in the shocked titanium Ti6Al4V. However, the variations in the mass density and the particle velocity are maximum in the shocked titanium Ti6Al4V but minimum in the shocked OFHC copper. This behaviour of mechanical quantities, especially, for the OFHC copper differs greatly from the titanium Ti6Al4V. Thus, the pressure, the temperature, the speed of sound, and the adiabatic bulk modulus increase with an increase in the Gr\"uneisen parameter $\Gamma_o$; however, the mass density and particle velocity decrease. It is also obvious from this figure that the jumps in the pressure and temperature are from lower to higher values of $\Gamma_o$ which shows that a shock wave leads to compression and heating of the material at the expense of streamwise velocity. 

\begin{figure}   
   \begin{minipage}{0.4\textwidth}
     \centering
     \includegraphics[width=1.5\linewidth, trim=0 .1cm .1cm 0, clip]{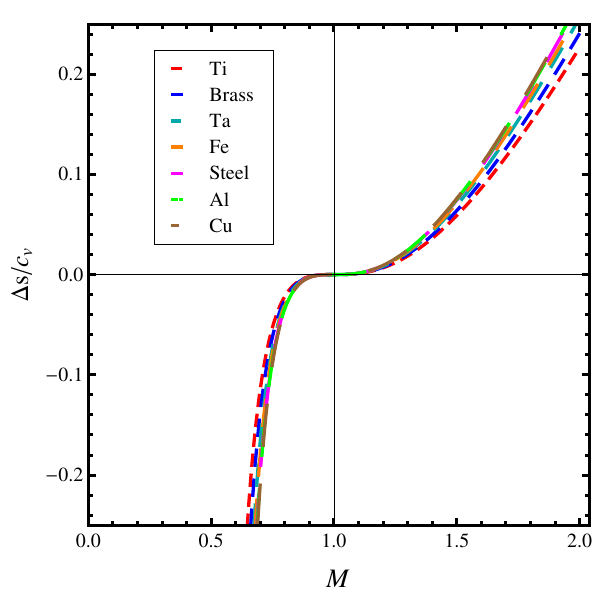}
       \end{minipage}\hfill
          \caption{Graphs of $\Delta s/c_v$ vs. $M$ with $G_m$=0.65.}\label{fig_3}
\end{figure}

Figure \ref{fig_3} shows the variation of change-in-entropy $\Delta s/c_v$ across the shock in titanium Ti6Al4V, brass(66\% copper and 34\% zinc), tantalum, iron, stainless steel 304, aluminum 6061-T6, and OFHC copper versus the upstream Mach number $M$ with $G_m$=0.65. It reveals that the change in entropy across the shock front increases with increasing values of $M$ and $\Gamma_o$. The change in entropy is maximum for OFHC copper and minimum for titanium Ti6Al4V. Figure \ref{fig_3} also illustrates that the change-in-entropy is very small when $M$ is close to unity and is negative when $M<1$. Thus, the shock waves do not occur in the condensed materials unless $M>1$, which is also in agreement with the second law of thermodynamics.

In the case of very weak shock waves, the properties of the material modify slightly across the shock. One can write for weak shocks $M=1+\varepsilon$, where $\varepsilon$ is a very small parameter, i.e., $\varepsilon\ll1$. Thus, RH shock jump relations (\ref{eq-3})-(\ref{eq-4}) in the present case are given by the following set of relations: 
\begin{eqnarray}\label{eq-6}
\frac{p}{p_{o}}=1+\frac{4(\Gamma+1)\varepsilon}{\Gamma+2}, \frac{\rho}{\rho_{o}}=1+\frac{4\varepsilon}{\Gamma+2}, \frac{u}{a_{o}}=\frac{4\varepsilon}{\Gamma+2}, \frac{T}{T_{o}}=\frac{\Gamma_{o}}{\Gamma}+\frac{4\Gamma_{o}\varepsilon}{\Gamma+2}, \nonumber \\
\frac{a}{a_{o}}=\left[\frac{\Gamma+1}{\Gamma_{o}+1}\left(1+\frac{4\Gamma\varepsilon}{\Gamma+2}\right)\right] ^{1/2}, \frac{K_s}{p_o}=(\Gamma+1)\left[1+\frac{4(\Gamma
+1)\varepsilon}{\Gamma+2}\right], \nonumber \\
\frac{\Delta s}{c_v}=ln\left[1+\frac{4(\Gamma+1)\varepsilon}{\Gamma+2}\right]-(\Gamma+1)ln\left[1+\frac{4\varepsilon}{\Gamma+2}\right]. \nonumber
\end{eqnarray} 
In view of these relations for weak shock waves, Anand \cite{Anand2022} has presented the shock dynamics for weak converging shock waves in solid materials.

In the limiting case of strong shock waves, the parameter $M$ is large i.e. $M\geq1$. Therefore, we may write $U\geq a_o$. Under this condition, the RH shock jump relations (\ref{eq-3})-(\ref{eq-4}) reduce to the following set of relations: 
\begin{eqnarray}\label{eq-5}
p=\frac{2\rho_o(\Gamma+1)}{(\Gamma_o+1)(\Gamma+2)}U^2, \rho=\rho_o\frac{\Gamma+2}{\Gamma}, u=\frac{2}{\Gamma+2}U, T=\frac{2\Gamma_o(\Gamma+1)T_o}{(\Gamma+2)^2}U^2, \nonumber \\
a=\frac{(\Gamma+1)}{(\Gamma+2)}\sqrt{\frac{2\Gamma}{\Gamma_o+1}}U, K_s=\frac{2\rho_o(\Gamma+1)^2}{(\Gamma_o+1)(\Gamma+2)}U^2, \nonumber \\
\frac{\Delta s}{c_v}=ln\left[\frac{2(\Gamma+1}{\Gamma+2}\left(\frac{U}{a_o}\right)^2\right]-(\Gamma+1)ln\left[\frac{\Gamma+2}{\Gamma}\right]. \nonumber
\end{eqnarray} 

It is notable that the increase in the pressure $p/p_o$, the temperature $T/T_o$, the adiabatic bulk modulus $K_s/p_o$, and the change-in-entropy $\Delta s/c_v$ can be infinitely large for sufficiently large shock strengths, i.e., $M\gg1$, but the increase in the downstream Mach number $M_f$, the mass density $\rho/\rho_o$, the particle velocity $u/a_o$, and the speed of sound $a/a_o$  are limited to $\frac{\sqrt{\Gamma(\Gamma_o+1)/2}}{\Gamma+1}$, $\frac{\Gamma+2}{\Gamma}$, $\frac{2}{\Gamma+2}$, and $\frac{(\Gamma+1)\sqrt{2\Gamma/(\Gamma_o+1)}}{\Gamma+2}$, respectively (see equations (\ref{eq-2})-(\ref{eq-4})). Recently, using the above shock jump relations, Anand \cite{Anand2025} has studied the convergence of strong cylindrical and spherical shock waves in solid materials.

\section{Conclusions}
\label{concl}
In this article, the RH conditions at the shock front have been proposed for condensed materials, and the effects on the mechanical quantities such as mass density, pressure, temperature, entropy, etc., of the shock-compressed OFHC copper, aluminum 6061-T6, titanium Ti6Al4V, stainless steel 304, brass (66\% copper and 34\% zinc), iron, and tantalum have been investigated. It is found that the material quantities vary with the strength of the shock. The variations in the pressure, the temperature, the speed of sound, the adiabatic bulk modulus, and the change in entropy are maximum for OFHC copper; however, these variations are minimum for titanium Ti6Al4V. Conversely, the variations in the particle velocity and the mass density are maximum for titanium Ti6Al4V, whereas minimum for OFHC copper. 

Thus, the present article disclosed the behaviour of condensed phase materials such as titanium Ti6Al4V, brass (66\% copper and 34\% zinc), tantalum, iron, stainless steel 304, aluminum 6061-T6, and OFHC copper. The handy forms of shock jump relations may be applied for investigating the weak as well as strong shock waves in solids, aerospace materials, etc. The findings of the present study might also prove to be a valuable reference for the physicists, geophysicists, material scientists, and engineers.

\end{document}